# Enabling optical steganography, data storage, and encryption with plasmonic colors

*Maowen Song,[1,2,†] Di Wang,[1,†] Zhaxylyk A. Kudyshev,[1,3] Yi Xuan,[1] Zhuoxian Wang,[1]*

*Alexandra Boltasseva,[1] Vladimir M. Shalaev,[1,*] and Alexander V. Kildishev[1,*]*

[1]Birck Nanotechnology Center, School of Electrical & Computer Engineering, Purdue

University, West Lafayette, Indiana 47907, United States

[2]National Laboratory of Solid-State Microstructures College of Engineering and Applied

Sciences and Collaborative Innovation Center of Advanced Microstructures,

Nanjing University Nanjing 210093, China.

[3]Center for Science of Information, Purdue University, West Lafayette, Indiana, 47907, United

States

[*]Correspondence and requests for materials should be addressed to V.M.S. and A.V.K.

E-mail: shalaev@purdue.edu, kildishev@purdue.edu

**Abstract**

Plasmonic color generation utilizing ultra-thin metasurfaces as well as metallic nanoparticles hold a great promise for a wide range of applications including color displays, data storage and information encryption due to its high spatial resolution and mechanical/chemical stability. Most of the recently demonstrated systems generate static colors; however, more advanced applications such as data storage require fast and flexible means to tune the plasmonic colors, while keeping them vibrant and stable. Here, a surface-relief aluminum metasurface that reflect polarization-tunable plasmonic colors is designed and experimentally demonstrated. Excitation of localized surface plasmons encodes discrete combinations of the incident and reflected polarized light into diverse colors. A single storage unit, namely a nanopixel, stores multiple-bit information in the orientation of its constituent nanoantennae, which is conveniently retrived by inspecting the reflected color sequence with two linear polarizers. Owing to the broad color variability and high spatial resolution of our metasurface, the proposed encoding approach holds a strong promise for rapid parallel read-out and encryption of high-density optical data. Our method also enables robust generation of dynamic kaleidoscopic images without the "cross-talk" effect. The approach opens up a new route for advanced dynamic steganography, high-density parallel-access optical data storage, and optical information encryption.

## 1. Introduction

In recent years, nanopatterned metals have been used to generate fade-free, environmentally friendly and stable colors adjustable by physical dimensions and composition of the nanostructures[1-3] As alternatives in producing colors, the so-called plasmonic nanostructures outperform conventional organic dyes and pigments in spatial resolution[4], mechanical/chemical robustness[5, 6], and resistance to chromic degradation[7]. When the light of a select wavelength strikes metallic nanoparticles, the impinging electric field drives the free electrons in metal to resonantly oscillate inside the nanoparticles, which in turn tailors the transmitted and reflected light spectra. Based on such coupled light-electron excitation called surface plasmon resonance (SPR), numerous systems comprising metallic nanoparticles and nanostructures have been demonstrated to generate vibrant and stable plasmonic colors. For example, subwavelength nanohole arrays[8–10], metal-insulator-metal gratings[11–13], highly dense arrangements of nanorods[14] and all metallic nanostructures[15–17] showcased exciting routes towards producing colors using propagating-type surface plasmons polaritons (SPPs)[10,18], localized surface plasmons resonance (LSPR)[14], or their combination[17,19].

The resonant frequency of a plasmonic color system depends on its dielectric environment, material's optical properties, and geometry of its plasmonic elements. The majority of the previously demonstrated plasmonic color-producing structures generate only static colors using nanoantennae of fixed shapes and dimensions. Moreover, these structures usually employ noble metals such as gold and silver as their constituent plasmonic materials that are expensive, have challenging fabrication (especially in thin films) and are incompatible with high-throughput manufacturing. These limitations are largely hindering the use of the proposed systems in real-life applications. Recently, a new line of research into polarization-tunable color-producing structures has been launched due to the increasing demands for innovative photonic technologies such as switchable displays[20], optical cryptography[21] and camouflage[22]. In contrast to isotropic nanodisks and nanohole arrays, anisotropic nanopixels are used to selectively switch the filtered color information by the polarization state of the incident light. For example, elliptical resonators[23], nanoblocks[24], cross-shaped protrusions[20] and apertures[21,22] with uneven arm lengths, are employed to create polarization-dependent color generation. In the demonstrated systems, two-color information states are simultaneously encoded in each nanopixel - a feature resulting in active color display technology[20], stereoscopic imaging[23], and high-density microimage encoding[22]. However, challenges remain in (i) increasing the number of polarization-dependent information states (colors)[20], and (ii) suppressing the undesired mixing of the overlaid images between the different polarization states, or the "cross-talk" effect[23].

Here, our APM relies on the LSPR – resonant free-electron oscillations at the metal-dielectric interface when stimulated by light – to modify the reflection spectrum. Since the LSPR is very sensitive to the relative angle between the anisotropic nanoantenna orientation and the incident/reflected light polarization, the reflected spectrum (color) varies with the relative angle between the two. As a result, the information can be stored in the orientation states of the nanoantennae, and conveniently retrieved by their reflected colors. Compared to polarization-tunable color-producing structures working in the transmission mode[25], a reflective metasurface, which is free from anti-reflection complications and propagation-accumulated phase, can be made with smaller thicknesses while simultaneously exhibiting high reflection efficiency. Such features can be utilized to shrink the device footprint and increase the saturation and intensity of the filtered colors in a bright-field optical microscope, leading to clear and vibrant microprints.

In contrast to the plasmonic color-producing structures based on composite metallic resonators, the use of all-Al design not only simplifies fabrication process by forming connected metal structures but also reduces the manufacturing cost. Moreover, the proposed design is compatible with commercial nanoimprint process[26] and promises to bridge the gap between proof of concept demonstrations and large scale manufacturing. With respect to color generation performance, the generated colors' dependence on both incident and reflected polarizations could be highly beneficial for allowing multiple images to be encoded in the same area and revealed by unique polarization keys. Employing these advantages, such an APM provides a viable solution to the problem of increasing the number of information states as well as suppressing the "cross-talk" effect by judiciously design. In this work, we experimentally showcase a storage density that is 5% greater than the state-of-the-art Blu-ray disc with a predicted 143× readout speed enhancement. The storage density can be increased even further by employing machine learning for color or spectral discrimination, along with a readout system of higher spectral sensitivity. Furthermore, such color-based data storage is compatible with our new parallel-processing data readout scheme where the information stored in multiple storage units is acquired simultaneously, significantly surpassing modern sequential systems in operation speed. In optical steganography, we encode more vivid "cross-talk"-free color images in a common area than any other relevant work. In information encryption, our APM is for example capable of encrypting the entire English alphabet in a single ciphertext element, showing enhanced functionality vs comparable digit-based plasmonic/optical encryption.

## 2. Results

### 2.1 Theoretical fundamentals and numerical simulations

The proposed metasurface consists of periodic arrangements of rectangular-shaped Al nanoantennae attached to an optically thick Al film. This architecture is capable of converting the diagonally-oriented (45° with respect to the long axis of the nanoantenna) linearly-polarized (LP) visible light into other polarization states. Importantly, a complete 90° optical rotation angle can be readily achieved with high reflection efficiency and device compactness due to the LSPR (see Supporting Information Figure S1). Al is chosen here due to its broad-band plasmonic properties - it supports plasmon resonances in the wavelength range from the ultraviolet to the near-infrared[7]. Moreover, the lower cost of Al and its stability in air makes it more suitable for our goal than conventional plasmonic metals such as silver and gold. Compared with other optical data storage mechanisms such as light-induced structure deformation[27,28] or material phase-change[29], a passive

Al metasurface is more robust and has longer data storing lifetime[30] thus is more suited for data storage.

As shown in the lower inset of Figure 1a, the unit cell of the APM spans an area of 250 nm × 250 nm with the optimized length (*l*), width (*w*) and thickness (*h*) of the nanoantenna being, $l = 200$ nm, $w = 80$ nm, and $h = 70$ nm. The period of the nanoantenna array is kept subwavelength to avoid diffraction, which causes unwanted sharp peaks in the reflectance spectra and deteriorates the color saturation. The red curve in Figure 1a indicates the phase difference $\delta$ that exhibits a pronounced 180° drop from ~460 nm to 600 nm. The four insets in Figure 1a show the calculated elliptical polarization states of the reflected *E*-field at four select wavelengths, implying that the polarization state of the strongest reflected light is wavelength-dependent, a quintessential feature of the optical-rotation effect[31].

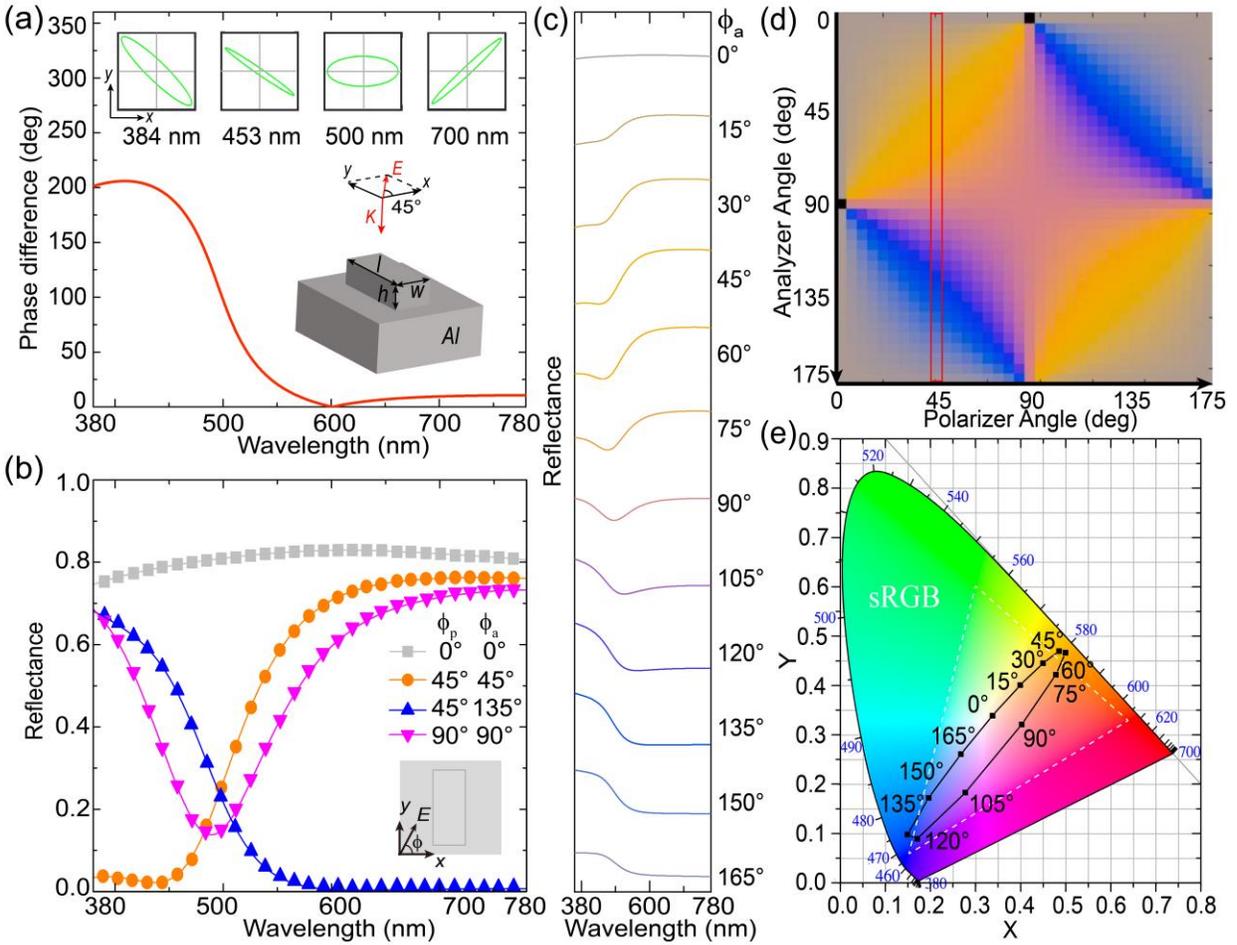

**Figure 1. Simulated results of the APM color filter**. **a,** Calculated reflected light phase difference between *x*- and *y*- polarizations. The four upper insets show the elliptical polarization states of the reflected wave at four select wavelengths. Lower inset: a schematic of a single unit cell containing an individual Al nanoantenna upon the incident polarization state. **b,** Simulated reflectance spectra corresponding to different combinations of polarizer-analyzer angles, leading to four distinct colors. **c,** Simulated spectra of light reflected from the APM when the analyzer is rotated from 0° to 165° with a step of 15°, while the polarizer is fixed at 45° with respect to the long axis of the nanoantenna. The lines in **c** are presented in colors calculated from a given

reflectance spectrum and the color-matching functions. **d, e,** Calculated color palette and corresponding color information on the CIE 1931 chromaticity diagram, when both the polarizer and analyzer are rotated. Colors are obtained with simulated reflectance spectra and color-matching functions. The elliptical color palette shown in **e** is remapped from the region enclosed by the red box in **d**.

The detailed simulation approach is given in the Methods section, and the polarized field calculations are described in Supporting Information. To make APM produce a wide palette of colors, various combinations of polarizer-analyzer angle differences ($\phi_p - \phi_a$) are utilized. Figure 1b shows the corresponding reflectance spectra at normal incidence. For ($\phi_p=0°$, $\phi_a=0°$), the reflectance spectrum becomes nearly flat across the entire visible spectrum, and the APM reflects a grey color. On the contrary, the spectra for ($\phi_p = 45°$, $\phi_a = 45°$), ($\phi_p = 90°$, $\phi_a = 90°$) and ($\phi_p = 45°$, $\phi_a = 135°$) show distinctive profiles, where the maximum reflectance exceeds 70% while the reflectance at off-resonant wavelengths is highly suppressed. Such a high contrast between the peak and dip values in the reflectance spectra improves color saturation and stability against fabrication uncertainties. Figure 1c depicts the simulated reflectance spectra when the analyzer is rotated from $\phi_a = 0°$ to $\phi_a = 165°$ with a step of 15° when the polarizer angle is fixed at 45° with respect to the long axis of the nanoantenna. Figure 1d shows the corresponding color palette versus polarizer-analyzer angle combinations. Based on the color matching functions defined by CIE, distribution of the colors in the CIE 1931 diagram can be clearly observed in Figure 1e.

## 2.2 Optical steganography: plasmonic "kaleidoscope"

We fabricate the simulated all-Al APM with standard electron-beam lithography (EBL), metallization and lift-off technique. A detailed description of the fabrication process is introduced in the Methods section. Figure 2a shows the measured reflectance spectra under the same polarizer-analyzer combinations as shown in Figure 1b. The excellent uniformity and high-fidelity profile of the fabricated nanoantennae lead to a good agreement between measured and simulated spectra. Some differences are due to the shape distortion at the fabricated nanoantenna corners, as well as changes in optical properties of nanostructured Al, which are not accounted for in simulations. To explore the potential use of this APM as a tunable plasmonic color filter, we photographed the colors with a chromatic charge-coupled device (CCD) camera integrated into an optical microscope. The detailed optical setup is schematically presented in Supporting Information Figure S2. For comparison, we present the simulated and experimentally photographed colors (blue, orange, magenta and grey) in the left and right columns of Figure 2b, respectively.

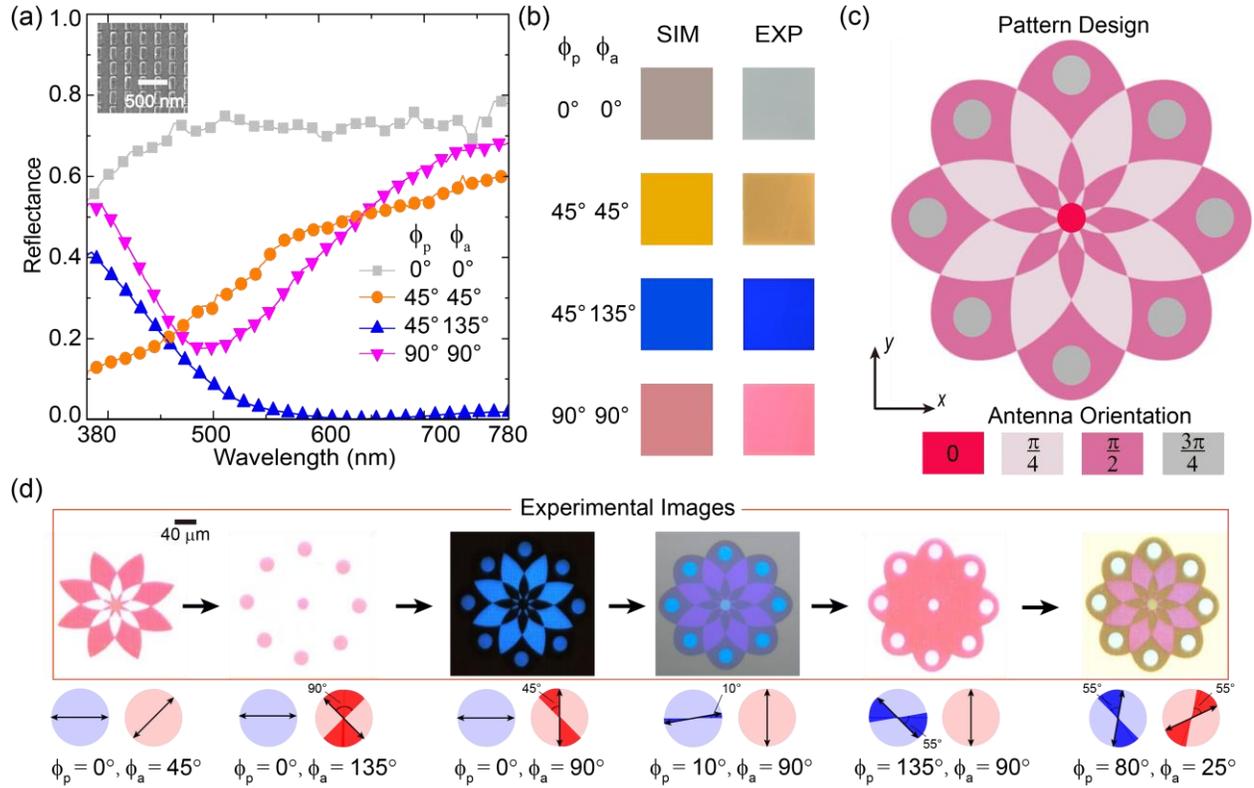

**Figure 2. Experimental results of the APM color filter and steganography. a,** Measured reflectance spectra of the APM under four different combinations of polarizer-analyzer angles corresponding to those in **Figure 1b**. Inset: SEM image of the Al nanoantennae taken in a random area of the fabricated APM. **b,** Comparison between the simulated colors (left) and CCD camera photographed colors (right) under the four polarizer-analyzer combinations (same as in **a**). All photographed images were obtained by collecting the reflected light into the CCD camera with a 20× objective lens (NA = 0.45) under white light illumination. **c,** Schematic of the steganographic flower pattern presented with different colors indicating different nanoantenna orientations. **d,** Experimental optical micrographs upon rotating polarizer and analyzer. The circles below the photographs represent the polarizer (blue) and analyzer (red) angles, with the highlighted regions

and corresponding numbers indicating the angles by which the polarizer/analyzer is rotated from the previous state.

The grey, blue, and magenta colors obtained (photographed) from the experiment agree reasonably well with the simulated ones, while the experimental orange color exhibits some discrepancy from the simulation. The discrepancy can be explained by the orange spectral curve depicted in Figure 2a – the broader resonance linewidth and relatively low reflectance profile lead to the reduced color saturation.

A broad palette of colors can be observed when rotating the polarizer and analyzer. In contrast to the dynamic plasmonic colors tuned by heat[32] or chemical reaction[33,34], the appearance of the color images encoded by our meticulously designed APM can be readily changed without causing any deformations to the structure. Such an APM opens up an avenue for advanced steganography – a technique used to conceal a message or image within another message or image. To demonstrate the concept of steganography with a plasmonic "kaleidoscope", we design a pattern of an eight-petal flower decorated with a core and eight circular speckles as depicted in Figure 2c. The areas occupied by differently oriented nanoantennae ($0, \frac{\pi}{4}, \frac{\pi}{2}, \frac{3\pi}{4}$, all nanoantenna orientations are with respect to the $x$-direction henceforth) are labelled by different colors and nanoantenna orientation angles. The square Al nanoantennae (with dimensions of $l \times w = 80$ nm $\times$ 80 nm) are used as the metasurface background, which blends in well with the peripheral part of the flower at specific polarization states. The performance of the plasmonic steganography is presented in Figure 2d. A magenta flower with eight petals is perfectly observed when the polarizer and analyzer are set at 0° and 45°. When the analyzer is rotated by 90° in the counter-clockwise (CCW) direction, the flower pattern undergoes dynamic profile change with the disappearance of the petals and emergence of "cluster of speckles", because the areas occupied by $\frac{\pi}{4}$ and $\frac{\pi}{2}$ oriented nanoantennae perfectly match the colors generated from the background square-shaped nanoantennae. Simultaneously, the areas occupied by 0 and $\frac{3\pi}{4}$ oriented nanoantennae render a clear-cut magenta flower which sharply contrasts with the background. Based on the same principle, more distinct images from this pattern can be revealed under many other polarizer and analyzer angle combinations (see Figure 2d), largely increasing the information capacity of the steganography technique.

**2.3 Towards a higher spatial resolution**

On top of expanding the available color palette, tuneability and abating the "cross-talk" effect, shrinking the dimensions of distinguishable nanopixels is of great importance for increasing the spatial resolution and information capacity of the APM. In order to find out the minimum nanopixel size that can support distinguishable colors, we fabricate checkerboard patterns with alternating nanopixels formed by nanoantennae in different orientations, as shown in Figs. 3a-c. When the polarizer is 0°, and the analyzer is 90° with respect to the *x*-direction, the alternating blue-black checkerboard pattern is distinctly observed as depicted in Figs. 3a,b.

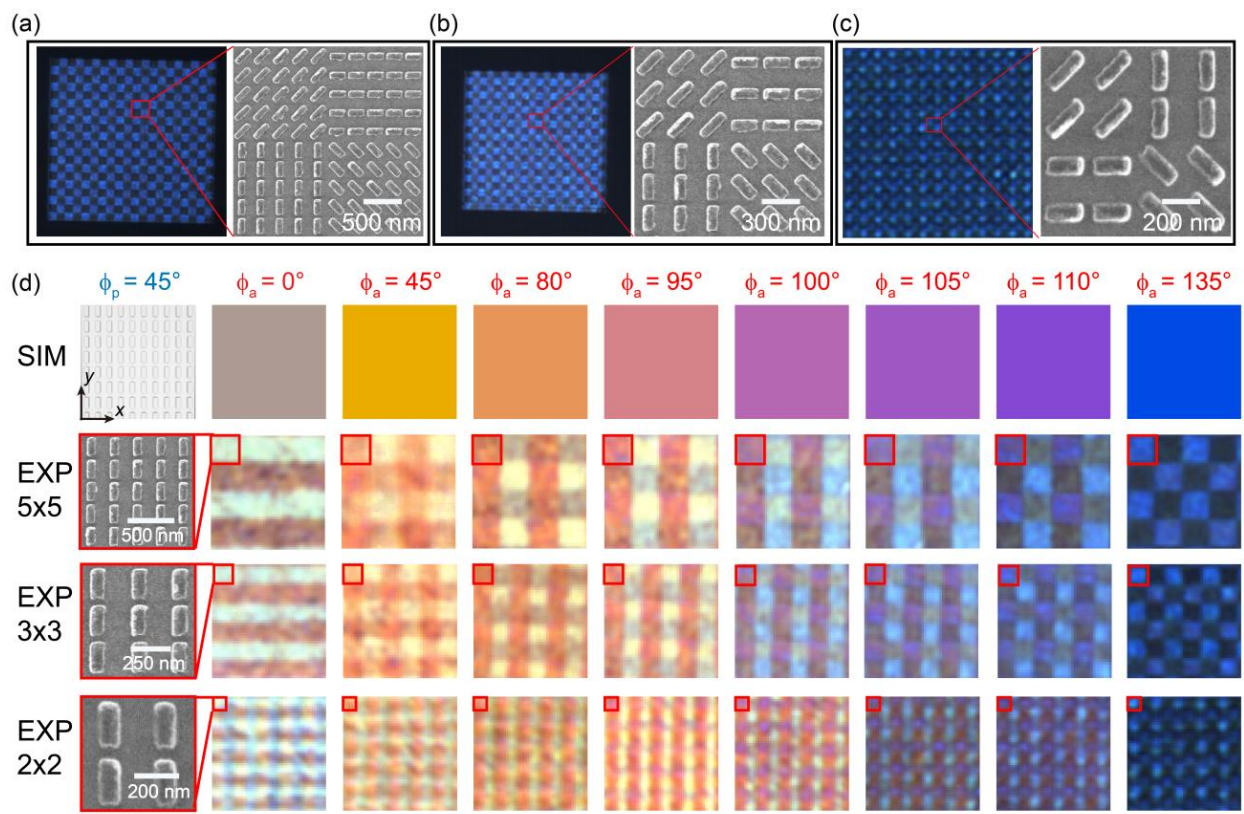

**Figure 3. Rendering colors by nanopixels**. **a-c**, The bright-field optical micrographs of the checkerboard patterned metasurface and the enlarged SEM images in the region of the red box. Each square red box in the checkerboard has a side of **a,** 1.25 µm **b,** 0.75 µm and **c,** 0.5 µm and consists of an array of **a**, 5×5 **b**, 3×3 and **c,** 2×2 nanoantenna unit cells. **d**, Simulated colors from infinite nanoantenna arrays (row 1), photographed images from 5×5 nanoantennae (row 2), 3×3 nanoantennae (row 3) and 2×2 nanoantennae (row 4). The corresponding analyzer angles are marked above; the polarizer angle is fixed at 45°.

Remarkably, Figure 3c shows that nanopixels with only 2 × 2 nanoantennae can exhibit vibrant alternating black-and-blue dots. The total area of such a nanopixel is 500 × 500 nm$^2$, which is equivalent to a spatial resolution of around 50,000 dots per inch (dpi) and exceeds the reported values in the most recent literature[35–40]. This effect originates from the LSPR's high field confinement at the nanoantenna/air interface[41–45]. By rotating the analyzer but fixing the polarizer at 45° with respect to the *x*-direction, more colors are expected to reflect from each nanopixel according to previous analysis. Interestingly, Figure 3d shows that eight different colors (see the regions enclosed by the red boxes) are observed by photographing the checkerboard pattern with a 100× and 0.9 NA objective. These images agree well with the simulated colors, implying that such an APM exhibits significant viewing angle tolerance and immense data storage potential with 500 × 500 nm$^2$ nanopixels.

**2.4 High-density data storage**

Owing to the high spatial resolution and great diversity in color generation, our APM can be used to augment current data storage technology. In conventional data storage devices, a storage unit accommodates only a single bit of information (0 or 1). With our APM, however, information is stored in the *orientation* of the nanoantennae and is retrieved by a set at of analyzers acquiring color sequences that uniquely match the nanoantenna orientations. Since a large variety of distinguishable colors can be rendered by rotating the nanoantennae, the amount of information stored in one APM storage unit can greatly surpass a single bit. In our proof-of-concept demonstration, we utilize eight nanoantenna orientations evenly distributed between 0 and $\frac{7\pi}{8}$ to represent eight distinct information states, which can be regarded as three bits of information with each nanoantenna orientation representing 000, 001, 010, 011, 100, 101, 110, or 111. Consequently, an APM can be programmed by EBL with nanopixels being the fundamental storage units each carrying 3-bit information. From the experimental result of the previous section, the nanopixels can be as small as 500 × 500 nm$^2$ (2 × 2 nanoantennae, or unit cells) in order to render distinguishable neighboring colors. Then, the storage density of the proposed APM (3 bits per 500 × 500 nm$^2$) is calculated to be 5% larger than that of the state-of-the-art Blu-ray technology (~1 bit per 320 × 274 nm$^2$, the dimensions of a Blu-ray bit are estimated from[46]). An example of such data-storage APM is shown in Figure 4a.

To translate the APM's antenna orientations into binary information, we propose a set of color codes that link the nanoantenna-rendered colors with the designated binary states, as well as a parallel-processing data readout system. Figure 4b presents the experimentally-obtained color codes (see more details on the color code generation in Supporting Information Figure S3). The figure shows that a given nanoantenna orientation renders a unique color sequence when sequentially imaged with analyzer angles of 0°, 45°, 90° and 135° (with the polarizer fixed at 45°). The core idea here is that a 3-bit code is assigned to each nanoantenna orientation state and then retrieved from the corresponding color sequence with an imaging system.

The readout system consists of four white-light sources and four CCD cameras as shown in Figure 4c. All four white-light sources are linearly polarized to 45° and illuminate adjacent regions on the pre-programmed APM. Each illuminated region hereafter referred to as a *frame*, may contain multiple nanopixels. For example, the frame shown in Figure 4a contains 16 nanopixels. CCD cameras record the rendered colors from the frame (each nanopixel in a frame may render a different color) after passing through four respective analyzers at angles mentioned above, and a local cache stores the color information on each nanopixel at each analyzer. Next time, when the array of frames moves forward by the length of one frame, the given frame is imaged at the next analyzer angle. When the frame passes all four CCD cameras, its constituent nanoantenna orientations, thereby stored data, is retrieved by looking up their color codes – the cache-stored color sequence. The four CCD images in Figure 4c are the color maps of the *single* frame shown in Figure 4a (also highlighted by the red box in Figure 4c), imaged under the four different analyzers.

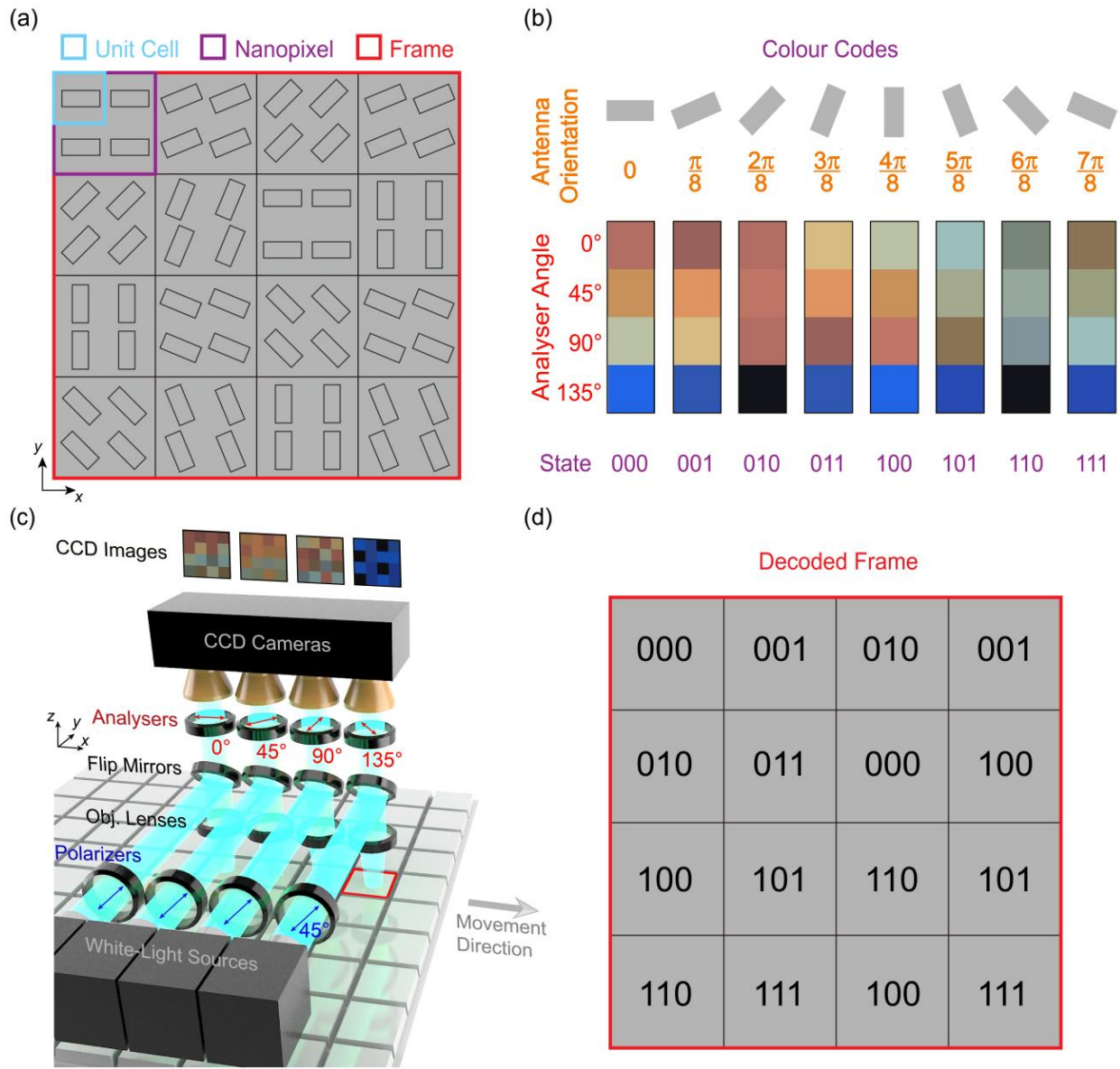

**Figure 4. APM as a data storage device. a,** A frame (red box) of a data-storage APM contains 16 nanopixels (the purple box highlights *one* nanopixel), each consisting of 4 unit cells (the blue box highlights *one* unit cell). A *nanopixel* serves as an indivisible data storage unit, which accommodates 3 bits of information. **b,** A unique color code is created for each nanoantenna orientation, which is then assigned a 3-bit information state, so that a piece of 3-bit information

can be stored in a nanopixel. **c,** The schematic of a proposed imaging system for fast APM information readout, which consists of four white-light sources, four CCD cameras and a sample-moving stage. The region highlighted by the red box shows the APM frame in **a**, and the four CCD images show the color maps of this frame imaged under the four respective analyzers. **d,** The stored binary data in **a** is retrieved after comparing the color sequence on each nanopixel and the color codes in **c**.

Consequently, the color sequence on each nanopixel can be obtained to retrieve the binary data stored in the frame, as shown in Figure 4d. The detailed data retrieval procedure is presented in Supporting Information Figure S4; the same procedure is applied to all frames of the APM to read the complete information. The advantage of the proposed readout scheme becomes apparent when the readout speed is considered. It takes 4X steps and a rotating analyzer for a setup with one light source and one camera to complete reading the information in one APM, where X is the number of frames in the APM; whereas only X + 3 steps are needed with the proposed parallel-processing setup, which also eliminates the need for a rotating analyzer. As we show in Supporting Information Section S6, with ultra-high-speed cameras (e.g. iX Camera i-SPEED 726) and other Blu-ray-equivalent components, it is possible to achieve a data readout speed of 18.3 Gbits/s, 143 times higher than the Blu-ray technology. It is worth noting that in the proposed readout scheme, the illuminating light need not be focused on one nanopixel, but instead enhances the readout speed by covering multiple nanopixels, provided that the color on each nanopixel can be spatially resolved by the CCD cameras (in this case, at each analyzer angle the cache stores a map of color pixels each corresponding to a nanopixel, as shown by the CCD images in Figure 4c).

The four-analyzer scheme used in the retrieval protocol is redundant for a 3-bit APM. Theoretically, even two analyzer angles are already sufficient to generate color codes that uniquely correspond to a nanoantenna orientation (it can be observed from Figure 4b, e.g., 0° and 90° analyzers). However, the redundancy is beneficial here; it can be used for error correction, hence enabling more robust information retrieval. Below we demonstrate the use of the four-analyzer scheme with more advanced APMs beyond 3 bits per nanopixel. Indeed, the data storage density in our APM can be further increased and is ultimately restricted by the CCD spectral resolution. We choose eight nanoantenna orientations in our experiment because the colors generated from these nanoantennae are easily distinguishable by the naked eye. Nonetheless, we show with simulation (see Supporting Information Figure S5) that 4-bit information can be reliably stored in a single nanopixel using 16 different orientation states ranging from 0 to $\frac{15\pi}{16}$. In this case, the storage density increases to 40% higher than a conventional Blu-ray disc, and the readout speed increases on average up to 191 times faster. As shown in this example, to increase the storage capacity per nanopixel in the APM, one needs only to utilize more nanoantenna orientations while keeping the nanoantenna geometry and the nanopixel size unchanged. In contrast to this, for the structures built on changing topologies (see e.g., ref[47]), the topolgical complexity significantly increases with storage density, and the structure needs to be enlarged to compensate for fabrication limitations. We may comment that implementing machine learning for color or spectral recognition[47] along with multiplayer storage system[48] are also viable steps towards further increasing the storage density while maintaining high readout accuracy. The potential in enhancing the data storage capacity and readout speed offered by our APM is a critical advancement both in the areas of plasmonic color generation and optical data storage technology.

**2.5 Optical information encryption**

The checkerboard pattern shown in the SEM image of Figure 3a can also be used in information encryption applications. In this context, the structure with four alternating nanoantenna orientations is regarded as a ciphertext. As shown in Figure 5a, when the ciphertext is imaged under 26 different polarizer-analyzer combinations, distinct color patterns can be obtained to represent the entire English alphabet. In order to attain the large variety of color patterns, the polarizer and analyzer angles are no longer limited to integer multiples of 45° but are chosen to optimize the distinguishability between patterns, and are used as keys to decrypt the information. As an example, in Figure 5b we show that when the ciphertext and two different key sets are sent to two recipients, different color patterns are perceived, from which one reads YES whereas the other reads OUT.

Although they are based on the same APM design, our proposed data storage and information encryption schemes are fundamentally different. In a data storage device, an APM needs to be pre-programmed to possess various nanopixels, and an invariant system acquires the information. In an information encryption system, the ciphertext is invariant, and various decryption key sets reveal the information.

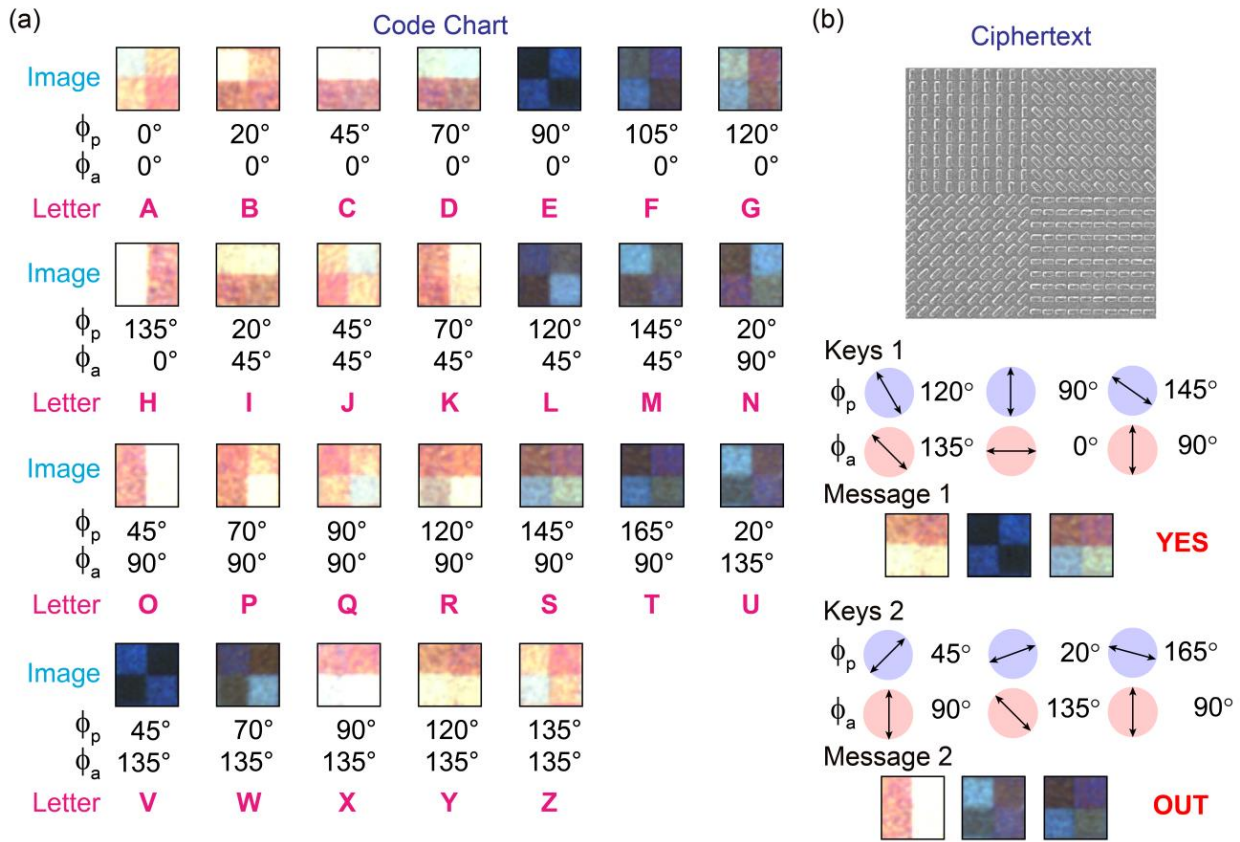

**Figure 5. APM as an information encryption device. a**, Code chart of the color patterns when a ciphertext APM is photographed with white light under 26 polarizer ($\phi_p$) and analyzer ($\phi_a$) combinations, each representing a letter in the English alphabet. The polarizer and analyzer angles are used as decryption keys. **b**, SEM of the ciphertext APM, and the different messages decrypted (YES and OUT) from the ciphertext by two sets of decryption keys.

## 3. Discussion

We propose a metasurface platform based on plasmonic color to realize three major functionalities in information technology – steganography, optical data storage, and encryption. With the metasurface, we demonstrate polarization-tunable colors in areas of $500 \times 500$ nm$^2$, presenting ~50,000 dpi resolution. The excellent color diversity leads to a judiciously designed optical steganography with vivid colors, abated "cross-talk" effect, and swift tuning. When used in optical data storage, such APM maintains the advantages of Blu-ray discs such as high durability (data are stored on physically-stable APM), low cost of operation (nearly no power consumption in idle state), and high data security/authenticity (stored data are immune to magnetic fields and cannot be altered). However, paired with the novel multiple-bit data storage scheme, our APM surpasses Blu-ray disc technology in both storage density and readout speed. Additionally, the metasurface is also used as a tool for information encryption, and multiple combinations of polarizer and analyzer angles are used as keys to decrypt the text phrases.

Besides the small dimensions of the pixel size, the fact that both polarizer and analyzer contribute to the optical response is essential. This feature not only presents the opportunity to erase, restore and tune the color image encoded in a common area but also improves the quality of imaging by abating the "cross-talk" effect which usually occurs in polarization-tunable color-producing structures[23]. Catalytic magnesium metasurface provides an alternative approach to tune the color[34], however, the chemical reaction slows down the tuning speed and the nested lithography limits the diversity of the information codes. In contrast to the known "five-dimensional" data storage technique[28], our metasurface can selectively generate different information states without causing any deformation to the nanoantennae, therefore is more robust for read-only data storage applications.

## 4. Conclusion

In summary, we design and experimentally demonstrate a versatile plasmonic color metasurface with a surface-relief Al film, which is used in advanced optical steganography, data storage, and encryption via tuning of reflected colors by rotating a polarizer and an analyzer. Employing the localised surface plasmon mode allows us to achieve a very high spatial resolution in the experiment. Based on the excellent color variability and high spatial resolution, we create and experimentally demonstrate an original technique to store multiple-bit information with the nanoantenna orientation states and retrieve this information using unique color codes. With this technique, the storage density already exceeds that of a conventional Blu-ray disc technology. Specially arranged analyzers enable a parallel data readout that outperforms traditional data readout system in operation speed. Also, we utilize the tunable colors to realize advanced optical steganography and encryption. The proposed approach offers a wide range of tunable colors without causing any deformation to the structure, thus demonstrating significant potential in advanced dynamic color applications.

## 5. Methods

**5.1 Numerical simulations**. The finite element method (FEM) based commercial software solver (CST Studio Suite 2017) is utilized to calculate the optical response of an elementary structure of the metasurface, with periodic boundary conditions used along *x* and *y* directions. The Al dielectric function is obtained from the data measured by a variable-angle spectroscopic ellipsometer (J. A. Woollam Co., W-VASE) (see Supporting Information Figure S6).

**5.2 Chromaticity calculations.** The colors induced by different polarizer-analyzer combinations are calculated using the simulated reflectance spectrum and the color-matching functions defined by the International Commission on Illumination (CIE)[2].

**5.3 Characterisation.** The colors of the APM are characterised by using a bright field reflection microscope setup (Nikon ECLIPSE 80i microscope) illuminated by a white light source (Nikon HALOGEN 12V50W LV-LH50PC). A chromatic CCD Camera (QIMAGING MicroPublisher 5.0 RTV) is used to photograph the color images. The reflectance spectra are measured using the variable-angle spectroscopic ellipsometer (J. A. Woollam Co., W-VASE). Figure 2(a) depicts the spectra obtained from a series of twenty measurements; the standard deviation at each point does not exceed the range of $\pm\, 2.1\times10^{-3}$. The light source is a xenon lamp with a broadband optical spectrum. The incident beam is focused to a spot with a diameter of 500 μm. The beam sequentially passes through a polarizer and then illuminates the sample. The incident angle is 18°. The overall area of the fabricated nanoantenna arrays used to obtain the results in Figure 2a is 500 × 500 µm$^2$; each array contains 2000 × 2000 unit cells.

**5.4 Fabrication of the metasurface.** The fabrication starts with electron beam evaporation of 100 nm thick aluminum (Al) film with 5 nm titanium (Ti) as an adhesion layer on a float glass (SiO$_2$) substrate, and spin coating of ~200 nm thick Poly(methyl methacrylate) (PMMA) on the Al film. The geometry of the nanoantennae is then defined on the PMMA by electron-beam lithography (EBL, JEOL JBX-8100FS) and development in a 1:3 methyl isobutyl ketone: isopropanol (MIBK:IPA) solution. Subsequently, 3 nm Ti and 70 nm-thick Al are deposited on the developed PMMA using electron beam evaporation. Finally, liftoff in heated acetone (70 °C) is carried out to define the Al nanoantennae on the optically thick Al film.

**Data availability**

The authors declare that all the data supporting the findings of this study are available within the paper and its Supporting Information.

**Author contribution**

M.S. and D.W. conceived and designed experiments, M.S. performed all the numerical simulations and imaging experiments, D.W. fabricated all the samples. M.S. conceived the color multiplexing concept. D.W., Z.A.K., and A.V.K. conceived the data storage method; A.V.K. conceived the parallel processing with fast readout. Y.X. provided help with the fabrication. Z.W. provided help with the measurement. M.S., D.W., and A.V.K. wrote the manuscript. A.B.,

V.M.S. and A.V.K. supervised the project, discussed the progress and results, and edited the manuscript. †These authors contributed equally to this work.

**Notes**

The authors declare no competing financial interest.

**Supporting Information**

Supporting Information is available from the Wiley Online Library or from the author.

**Acknowledgment**

M.S. would like to acknowledge the Chinese Scholarship Council (CSC, No. 201606050044) for financial support. Purdue co-authors would like to acknowledge support for optical characterization and fabrication from the U.S. Office of Naval Rsearch Grant (N00014-18-1-2481) and numerical modelling from DARPA/DSO Extreme Optics and Imaging (EXTREME) program (HR00111720032).

**References**

1. Song, M. *et al.* Achieving full-color generation with polarization-tunable perfect light absorption. *Opt. Mater. Express* **9**, 779 (2019).

2. Song, M. *et al.* Colors with plasmonic nanostructures: A full-spectrum review. *Applied Physics Reviews* **6,** 041308 (2019).

3. Tan, S. J. et al. Plasmonic Color Palettes for Photorealistic Printing with Aluminum Nanostructures. *Nano Lett.* **14**, 4023-4029 (2014).

4. Kumar, K. *et al.* Printing color at the optical diffraction limit. *Nat. Nanotechnol.* **7**, 557–561 (2012).

5. Clausen, J. S. *et al.* Plasmonic Metasurfaces for Coloration of Plastic Consumer Products. *Nano Lett.* **14**, 4499–4504 (2014).

6. Roberts, A. S., Pors, A., Albrektsen, O. & Bozhevolnyi, S.I. Subwavelength Plasmonic Color Printing Protected for Ambient Use. *Nano Lett.* **14**, 783-787 (2014).

7. James, T. D., Mulvaney, P. & Roberts, A. The Plasmonic Pixel: Large Area, Wide Gamut Color Reproduction Using Aluminum Nanostructures. *Nano Lett.* **16**, 3817-3823 (2016).

8. Yokogawa, S., Burgos, S. P. & Atwater, H. A. Plasmonic Color Filters for CMOS Image Sensor Applications. *Nano Lett.* **12,** 4349–4354 (2012).

9. Burgos, S. P., Yokogawa, S. & Atwater, H. A. Color Imaging via Nearest Neighbor Hole Coupling in Plasmonic Color Filters Integrated onto a Complementary Metal-Oxide Semiconductor Image Sensor. *ACS Nano* **7,** 10038–10047 (2013).

10. Barnes, W. L., Dereux, A. & Ebbesen, T. W. Surface plasmon subwavelength optics. *Nature* **424,** 824–830 (2003).

11. Xu, T., Wu, Y.-K., Luo, X. & Guo, L. J. Plasmonic nanoresonators for high-resolution color filtering and spectral imaging. *Nat. Commun.* **1,** 1–5 (2010).

12. Cai, W. *et al.* Metamagnetics with rainbow colors. *Opt. Express* **15,** 3333 (2007).

13. Wang, H. *et al.* Full Color Generation Using Silver Tandem Nanodisks. *ACS Nano* **11,** 4419–4427 (2017).

14. Si, G. *et al.* Reflective plasmonic color filters based on lithographically patterned silver nanorod arrays. *Nanoscale* **5,** 6243 (2013).

15. Zhang, J., Ou, J.-Y., MacDonald, K. F. & Zheludev, N. I. Optical response of plasmonic relief meta-surfaces. *J. Opt.* **14,** 114002 (2012).

16. Goh, X. M., Ng, R. J. H., Wang, S., Tan, S. J. & Yang, J. K. W. Comparative Study of Plasmonic Colors from All-Metal Structures of Posts and Pits. *ACS Photonics* **3,** 1000–1009 (2016).

17. Song, M. *et al.* Color display and encryption with a plasmonic polarizing metamirror. *Nanophotonics* **7,** 323–331 (2018).

18. Kristensen, A. et al. Plasmonic color generation. *Nature Reviews Materials* **2**, 16088 (2016).

19. Shrestha, V. R., Lee, S.-S., Kim, E.-S. & Choi, D.-Y. Aluminum Plasmonics Based Highly Transmissive Polarization-Independent Subtractive Color Filters Exploiting a Nanopatch Array. *Nano Lett.* **14,** 6672–6678 (2014).

20. Ellenbogen, T., Seo, K. & Crozier, K. B. Chromatic Plasmonic Polarizers for Active Visible Color Filtering and Polarimetry. *Nano Lett.* **12,** 1026–1031 (2012).

21. Li, Z., Clark, A. W. & Cooper, J. M. Dual Color Plasmonic Pixels Create a Polarization Controlled Nano Color Palette. *ACS Nano* **10,** 492–498 (2016).

22. Heydari, E. *et al.* Plasmonic Color Filters as Dual-State Nanopixels for High-Density Microimage Encoding. *Adv. Funct. Mater.* **27,** 1701866 (2017).

23. Goh, X. M. *et al.* Three-dimensional plasmonic stereoscopic prints in full color. *Nat. Commun.* **5,** 5361 (2014).

24. Nagasaki, Y., Suzuki, M. & Takahara, J. All-Dielectric Dual-Color Pixel with Subwavelength Resolution. *Nano Lett.* **17,** 7500–7506 (2017).

25. Bao, Y. *et al.* Full-color nanoprint-hologram synchronous metasurface with arbitrary hue-

saturation-brightness control. *Light Sci. & Appl.* **8**, 95 (2019).

26. Guo, L. J. Nanoimprint Lithography: Methods and Material Requirements. *Adv. Mater.* **19,** 495–513 (2007).

27. Gu, M., Li, X. & Cao, Y. Optical storage arrays: a perspective for future big data storage. *Light Sci. Appl.* **3**, e177–e177 (2014).

28. Zijlstra, P., Chon, J. W. M. & Gu, M., Five-dimensional optical recording mediated by surface plasmons in gold nanorods. *Nature* **459**, 410–413 (2009).

29. Feldmann, J. *et al.* Calculating with light using a chip-scale all-optical abacus. *Nat. Commun.* **8**, 1256 (2017).

30. Zhang, J., Gecevičius, M., Beresna, M. & Kazansky, P. G. Seemingly Unlimited Lifetime Data Storage in Nanostructured Glass. *Phys. Rev. Lett.* **112**, 033901 (2014).

31. Wu, S. *et al.* Enhanced Rotation of the Polarization of a Light Beam Transmitted through a Silver Film with an Array of Perforated S-Shaped Holes. *Phys. Rev. Lett.* **110**, 207401 (2013).

32. Zhu, X., Vannahme, C., Højlund-Nielsen, E., Mortensen, N. A. & Kristensen, A. Plasmonic color laser printing. *Nat. Nanotechnol.* **11**, 325–329 (2016).

33. Chen, Y. *et al.* Dynamic Color Displays Using Stepwise Cavity Resonators. *Nano Lett.* **17**,

5555–5560 (2017).

34. Duan, X., Kamin, S. & Liu, N. Dynamic plasmonic color display. *Nat. Commun.* **8**, 14606 (2017).

35. Sun, S. *et al.* All-Dielectric Full-Color Printing with TiO2 Metasurfaces. *ACS Nano* **11**, 4445–4452 (2017).

36. Proust, J., Bedu, F., Gallas, B., Ozerov, I. & Bonod, N. All-Dielectric Colored Metasurfaces with Silicon Mie Resonators. *ACS Nano* **10**, 7761–7767 (2016).

37. Kim, H. *et al.* Structural color printing using a magnetically tunable and lithographically fixable photonic crystal. *Nat. Photonics* **3**, 534 (2009).

38. Horie, Y. *et al.* Visible Wavelength Color Filters Using Dielectric Subwavelength Gratings for Backside-Illuminated CMOS Image Sensor Technologies. *Nano Lett.* **17**, 3159–3164 (2017).

39. Franklin, D., Frank, R., Wu, S.-T. & Chanda, D. Actively addressed single pixel full-color plasmonic display. *Nat. Commun.* **8**, 15209 (2017).

40. Xiong, K. *et al.* Plasmonic Metasurfaces with Conjugated Polymers for Flexible Electronic Paper in Color. *Adv. Mater.* **28**, 9956–9960 (2016).

41. Luo, X., Tsai, D., Gu, M. & Hong, M. Subwavelength interference of light on structured

surfaces. *Adv. Opt. Photonics* **10**, 757 (2018).

42. Luo, X., Tsai, D., Gu, M. & Hong, M. Extraordinary optical fields in nanostructures: from sub-diffraction-limited optics to sensing and energy conversion. *Chem. Soc. Rev.* **48**, 2458–2494 (2019).

43. Zang, X. *et al.* Polarization Encoded Color Image Embedded in a Dielectric Metasurface. *Adv. Mater.* **30**, 1707499 (2018).

44. Fang, J. *et al.* Enhanced Graphene Photodetector with Fractal Metasurface. *Nano Lett.* **17**, 57–62 (2017).

45. Song, M. *et al.* Nanofocusing beyond the near-field diffraction limit via plasmonic Fano resonance. *Nanoscale* **8,** 1635–1641 (2016).

46. *White Paper Blu-ray Disc $^{TM}$ Format*. (Blu-ray Disc $^{TM}$, 2015).

47. Wiecha, P. R., Lecestre, A., Mallet, N., Larrieu, G., Pushing the limits of optical information storage using deep learning. *Nature Nanotechnology* **14**, 237-244 (2019).

48. Taylor, A. B., Michaux, P., Mohsin, A. S. M. & Chon, J. W. M. Electron-beam lithography of plasmonic nanorod arrays for multilayered optical storage. *Opt. Express* **22**, 13234 (2014).

# Supporting Information

# Enabling optical steganography, data storage, and encryption with plasmonic colors

*Maowen Song,[1,2,†] Di Wang,[1,†] Zhaxylyk A. Kudyshev,[1,3] Yi Xuan,[1] Zhuoxian Wang,[1] Alexandra Boltasseva,[1] Vladimir M. Shalaev,[1,*] and Alexander V. Kildishev[1,*]*

[1]Birck Nanotechnology Center, School of Electrical & Computer Engineering, Purdue University, West Lafayette, Indiana 47907, United States

[2]National Laboratory of Solid-State Microstructures College of Engineering and Applied Sciences and Collaborative Innovation Center of Advanced Microstructures, Nanjing University Nanjing 210093, China.

[3]Center for Science of Information, Purdue University, West Lafayette, Indiana, 47907, United States

[*]Correspondence and requests for materials should be addressed to V.M.S. and A.V.K.

E-mail: shalaev@purdue.edu, kildishev@purdue.edu

**S1. Localized Surface Plasmon Resonance in Al Nanoantenna**

In this section, we discuss theoretically the localized surface plasmon resonance (LSPR) in aluminum (Al) nanoantennae. When a diagonally oriented linear-polarized (LP) light beam is normally incident on the anisotropic plasmonic metasurface

(APM), the electric field of the reflected light can be decomposed into the two eigenstates in *x*- and *y*-polarizations:

$$E_x = |E_x| \cos \tau$$
$$E_y = |E_y| \cos(\tau + \delta) = |E_y|(\cos \tau \cos \delta - \sin \tau \sin \delta) \tag{S1}$$

here $|E_x|$ and $|E_y|$ are the amplitudes of the *E*-field linearly polarized along the *x* and *y* directions, $\tau = \omega t - kz$, and $\varphi = |\varphi_x - \varphi_y|$ is a phase difference between the reflected *x*- and *y*- polarizations. $E_x$, $E_y$, $\delta_x$, and $\delta_y$ for a given unit cell design can be obtained using a commercial full-wave solver based on the finite element method (FEM). The following expression deduced from equations (S1) governs the relationship between the amplitude, phase difference $\delta$ and the elliptical polarization state of the reflected field:

$$\frac{E_x^2}{|E_x^2|} + \frac{E_y^2}{|E_y^2|} - 2\cos\frac{E_y E_x}{|E_y||E_x|} = \cos^2\tau - \cos(\tau - \delta)\cos(\tau + \delta) = \sin^2\varphi \tag{S2}$$

The polarization state and phase define the reflectance upon an arbitrary combination of polarizer-analyzer angles; the reflectance is calculated from ref[1]:

$$|E|^2 = \frac{1}{4}\begin{bmatrix}(E_x^2 + E_y^2)(1 + \cos 2\phi_a \cos 2\phi_p) + \\ (E_x^2 - E_y^2)(\cos 2\phi_a + \cos 2\phi_p) + \\ 2E_x E_y \sin\phi_a \sin\phi_p \cos\delta\end{bmatrix} \tag{S3}$$

Figure S1 depicts the analysis of reflectance spectra and resonance behavior depending on the nanoantenna dimensions. Figure S1A presents the reflectance spectra of the APM, where the length *L* of the nanoantenna (long axis) is changing from 150 nm to 230 nm by a constant step of 10 nm, while its width *w* (short axis) remains fixed at 80 nm. For the incident polarization along the major axis, the resonant dips of the reflectance spectra redshift with the increase of nanoantenna length (Figure S1B), while for the incident polarization along the minor axis, the reflectance spectra show negligible differences (Figure S1C). The length of the nanoantenna along the incident polarization direction determines the spectral position of the LSPR (see, for example, ref[2] and references within). To determine the length ($L$) of the nanoantenna for achieving resonance at a certain wavelength ($\lambda_{eff}$), we have adopted results presented in ref[2]:

$$L = \frac{\lambda_{eff}}{2n_{eff}} - 2\delta(\lambda_{eff}) \quad (S4)$$

here, $\delta$ is the evanescent extension of the resonant antenna mode and determined via fitting as $\delta = \delta_0 - \delta_1 \lambda_{eff} - \delta_2^2 \lambda_{eff}^2$ ($\delta_0 = 1.2\ \mu m$, $\delta_1 = -3.4\ nm^{-1}$, $\delta_2 = 0.0024\ nm^{-1}$). The effective permittivity of the local medium around the antenna is described as $\varepsilon_{eff}(\lambda) = \text{Re}[\varepsilon_{eff}(\lambda)] + i\text{Im}[\varepsilon_{eff}(\lambda)] = \frac{1}{2}[\varepsilon_{Al}(\lambda) + \varepsilon_{air}]$. Using the effective permittivity, the effective index can be directly obtained from $n_{eff}(\lambda) = \sqrt{\frac{1}{2}\big[|\varepsilon_{eff}(\lambda)| + \text{Re}[\varepsilon_{eff}(\lambda)]\big]}$. Figure S1B shows the length of the antenna as a function of resonant wavelength (lower inset shows the $\delta(\lambda_{eff})$ function). Figure S1D depicts the on-resonance electric and magnetic field distributions at the vertical cross-section of the nanoantenna with $L = 200$ nm and $\lambda = 500$ nm, where the fields are highly localized at the corners and edges of the nanoantenna, indicating the excitation of LSPR.

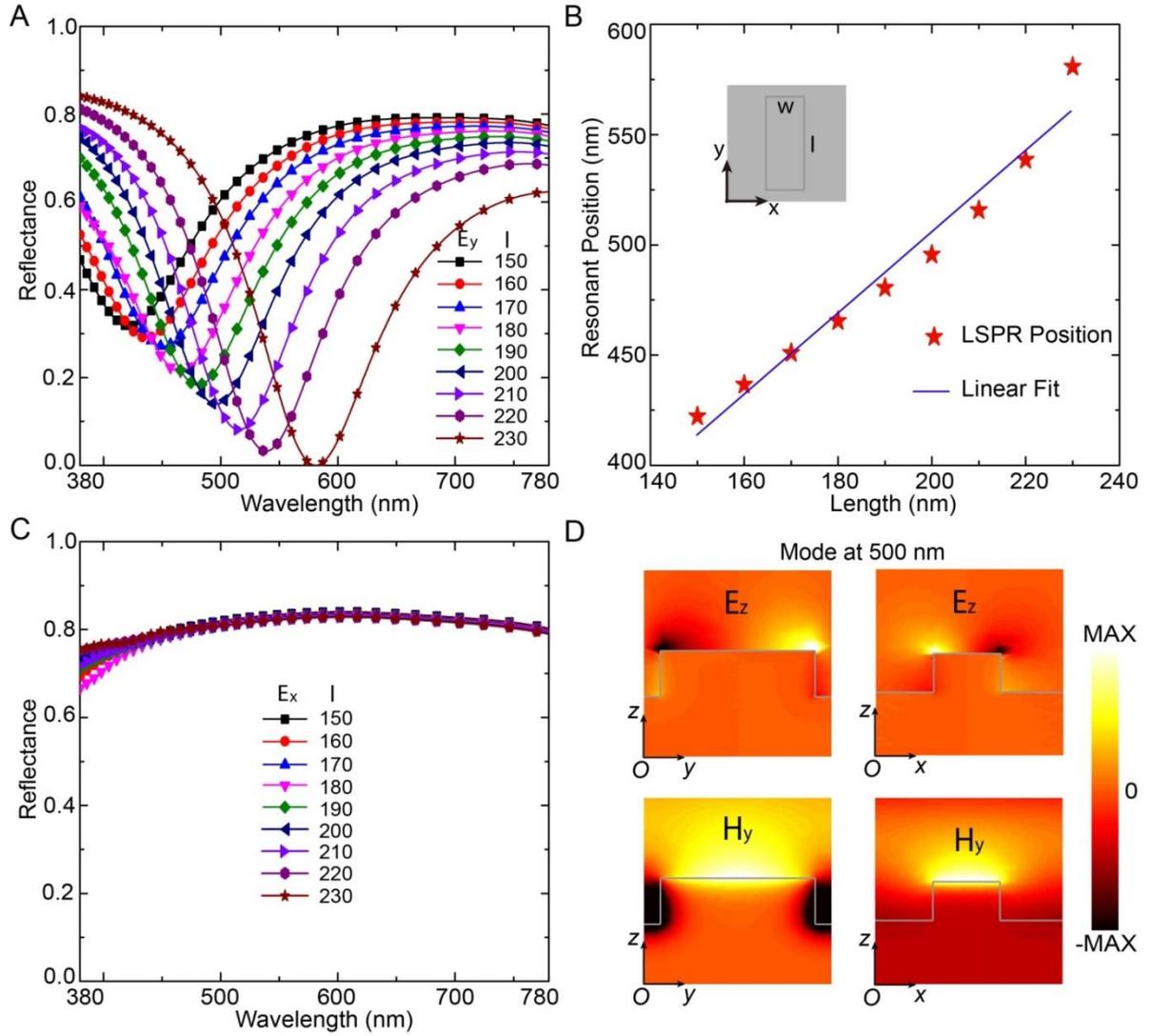

**Figure S1. Localized resonant modes analysis.** (**A**) Calculated reflectance spectra of the APM; the length l of the rectangular-shaped Al nanoantennae is changing from 150 nm to 230 nm with a step of 10 nm, while its width *w* is fixed at 80 nm. (**B**) Calculated positions of reflectance spectral dip versus the nanoantenna length. The blue line is a fit of the discrete LSPR positions. The upper inset shows the nanoantenna dimensions, and the lower inset shows the $\delta(\lambda_{eff})$ function. (**C**) Reflection spectra of the APM when the nanoantenna length *L* is varied from 150 nm to 230 nm under *x* polarized incidence. (**D**) Electric and magnetic field distributions of the resonant mode at a wavelength of 500 nm.

## S2. Experimental Setup for Photographing the APMs

Figure S2 illustrates the working principle of the APM. Optically broadband LP light is obtained from a halogen lamp with a broadband linear polarizer (Polarizer, Figure S2). A beam splitter guides the incident LP light to the APM, and the reflected light to a chromatic camera. Since the APM leads to the occurrence of dispersive optical-rotation effect (see the main text), a distinct reflectance spectrum is observed for each rotation state of the analyzer (Analyzer, Figure S2) – yet another broadband linear polarizer. In our experiment, we use the microscope Nikon Eclipse 80i, and LV-LH50PC HALOGEN 12V50W as the halogen lamp for illumination.

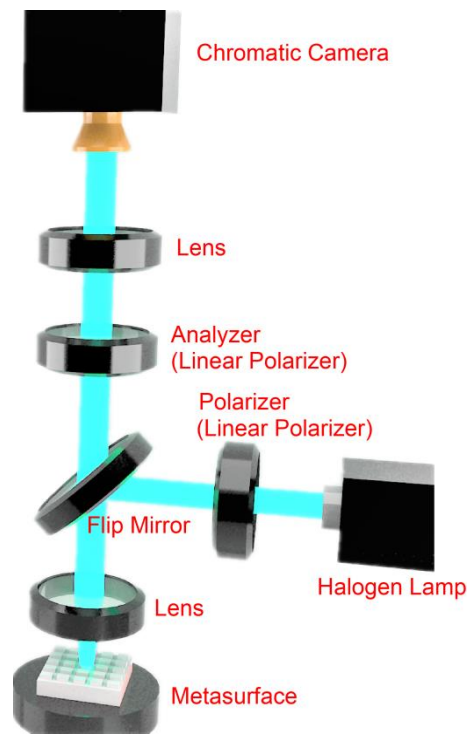

**Figure S2. Experiment setup.** Two broadband linear polarizers, Analyzer and Polarizer, which cover the visible region (380 nm – 780 nm), and a chromatic camera are integrated with a microscopic setup to image colors reflected from the APM under various Analyzer and Polarizer combinations. The halogen lamp is a broadband white light source.

### S3. Color Codes for Data Storage: Initial Experimental Validation

In this section, we present the method to generate the color codes (Figure 4B of the main text) that are used to retrieve information from the APM. We fabricate a wheel pattern containing the eight nanoantenna orientations and image the wheel using four analyzer rotation states (0°, 45°, 90°, 135°) with the polarizer rotation angle fixed at 45°.

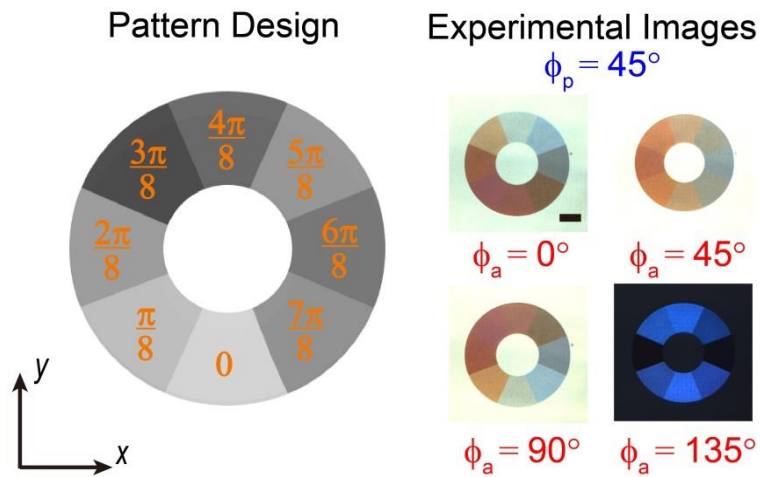

**Figure S3. Experimental color codes generation.** Left panel: The designed pattern is a wheel consisting of eight equal segments, each occupied by nanoantennae with the orientation marked by the orange-colored numbers. Right panel: the photographed images under four analyzer angles with polarizer fixed at 45°. The diverse colors from all eight segments under each analyzer angle are picked to construct the color codes in Figure 4B of the main text.

Figure S3 shows the wheel pattern design and the experimentally obtained images. Each nanoantenna orientation state renders a different color at a given analyzer rotation angle, thus allowing us to construct a color code sequence for each nanoantenna orientation examined with four analyzer states, as shown in Figure 4B of the main text. Assigning an individual 3-bit information state to each 4-color code, we, therefore, establish a unique correspondence between the nanoantenna orientation state and the 3-bit information. As discussed in the main text, the redundancy in analyzer angles is conducive to more robust data read-out and is suitable for more advanced APM data storage systems (see Section S5 of Supporting Information).

### S4. Reading APM-Stored Information

This section uses an example to illustrate how the binary information stored in a single APM nanopixel is read out using the setup in Figure 4C of the main text. A nanopixel with the nanoantenna orientation state of 22.5° is first imaged with 0°-analyzer – the camera records a maroon color (Figure S4A). As the sample moves, the sample nanopixel is then sequentially imaged with analyzers at 45°, 90°, and 135° rotation states, so that orange, beige, and blue colors are acquired by the camera, respectively (Figure S4, B to D). During the entire imaging process, the rotation state of polarizer is fixed at 45°. A local cache stores the maroon-orange-beige-blue 4-color sequence from the given nanopixel, and looks it up in the color codes (Figure 4B of the main text), to retrieve a binary code of 001 stored in this nanopixel. Several nanopixels in a frame are imaged at once; therefore, the read-out speed is significantly increased compared to the conventional single-point read-out systems (see Supporting Information Section S6).

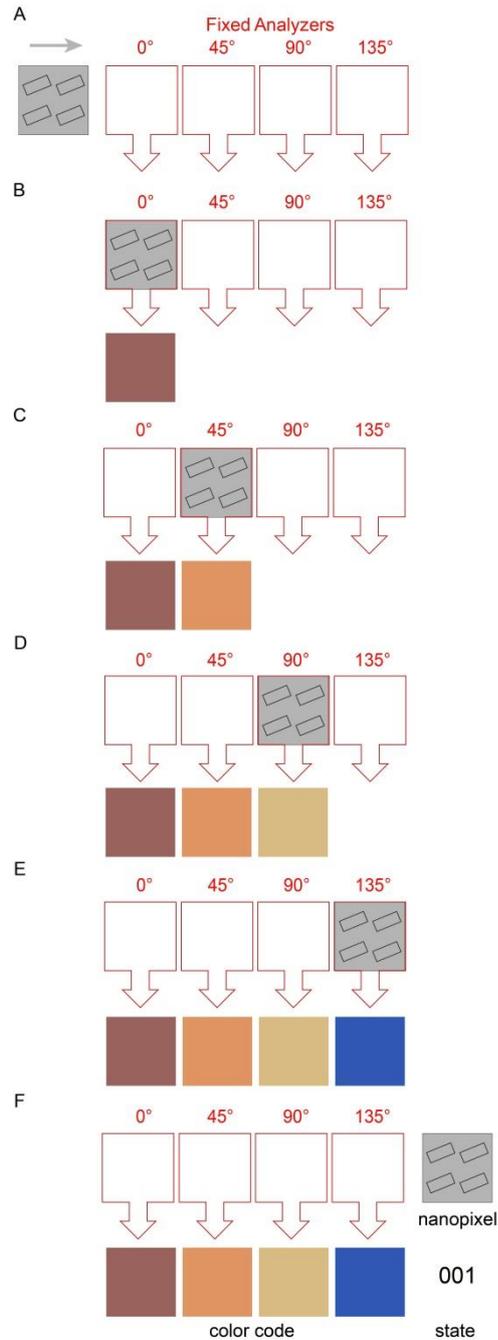

**Figure S4. Reading binary information stored in an APM nanopixel.** The nanopixel containing $\pi/8$-oriented nanoantennae is imaged with Analyzer at 0°, 45°, 90°, and 135° rotation states, respectively, and renders maroon **(A)**, orange **(B)**, beige **(C)**, and blue **(D)** colors. The color sequence in d is matched with the look-up table of Figure 4B in the main text, to retrieve the binary code 001.

## S5. Enhanced Data Storage Capacity

As stated in the main text, our APM can be used in data storage applications where the information is stored in the nanoantenna orientation states. In the main text, we demonstrate the concept using eight nanoantenna orientations experimentally – the photographed colors generated from these orientations are easily distinguishable even by the naked eye. Here, we show with a numerical simulation that the data storage capacity can be further enhanced by utilizing 16 different nanoantenna orientations, so that each orientation can represent a 4-bit word (a tetrad), thereby doubling the data storage capacity compared to the demonstration in the main text. Figure S5 shows the simulated color codes of the 16 nanoantenna orientation states obtained with the four analyzer rotation states (0°, 45°, 90°, 135°) at a fixed polarizer angle of 45°. It can be concluded that the 16-color sequences uniquely represent the nanoantenna orientation states, therefore, they can be used to retrieve the corresponding information states. It is to be noted, however, that as the data storage capacity increases, some of the color sequences become indiscernible by the naked eye. Improving the robust read-out may require (i) a higher spectral resolution of the camera, (ii) optimization of the angular states of both antennas and analyzer, and (iii) the use of the overdetermined number of the analyzer rotation states (as we already do for the 3-bit states in the main text).

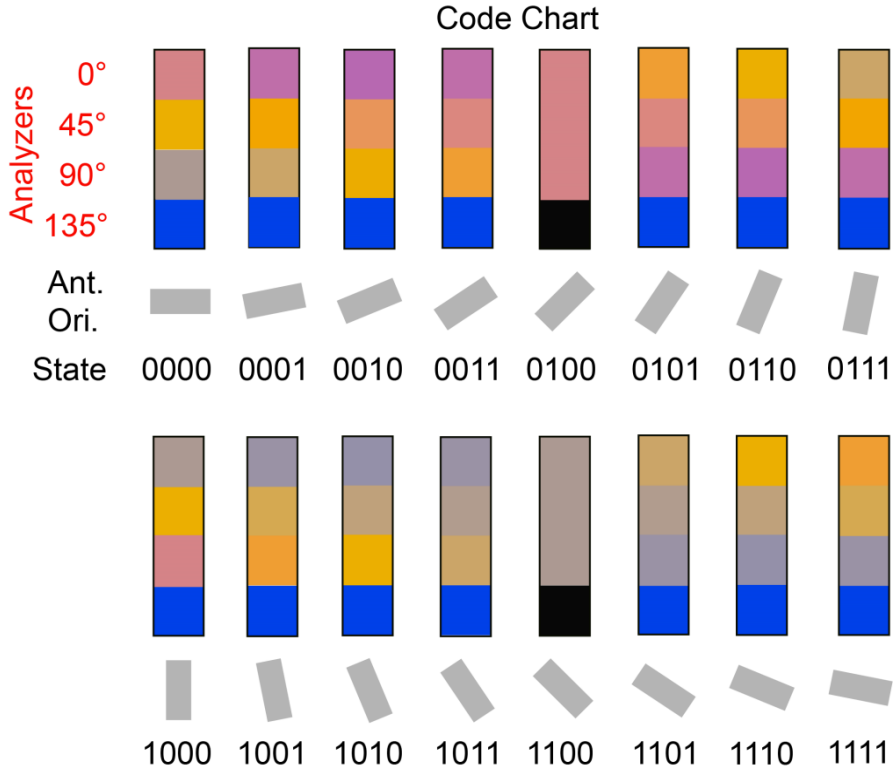

**Figure S5. Simulated 4-color sequences encoding 4 bits of information (16 states).** The nanoantenna orientation states change from 00 to 15 π/16 with a step of π/16. The code chart shows that the reflected 4-color sequences obtained with analyzers with four rotation states (0°, 45°, 90°, 135°) uniquely match the nanoantenna orientation states. The experimental setup is identical to the setup shown in Figure 4C of the main text.

### S6. The estimate of Data Readout Speed

In this section, we discuss the theoretical limit of the data read-out speed using the system shown in Figure 4 of the main text. In the following calculations, we consider a 3-bit APM, i.e. each nanopixel of area 500 × 500 nm$^2$ stores 3 bits of information. The data read-out system is an image-based parallel processing system, meaning that all the information in a frame (see main text for the definition of a frame) can be retrieved at once by taking an image of the frame. As a result, the factors to consider regarding the read-out speed are: 1. The number of bits per frame; 2. The time required for a camera to take an image of the frame; 3. The time required for the translational stage or disc spinner to move from one frame to another. In the following, we first calculate the three factors, respectively, and then obtain the read-out speed by combining them.

**1. The number of bits per frame:** The frame area is mainly limited by the field of view of the objective lens. In our experiment, we use an objective lens with a field of view (diameter) of 0.18 mm, which means the maximum frame area is

$$A_{frame} = \pi \left(\frac{1}{2} D_{frame}\right)^2 = \pi \left(\frac{1}{2} 0.18\right)^2 = 2.54 \times 10^{-8} m^2 \qquad (S5)$$

and the number of nanopixels in each frame is $N = \frac{A_{frame}}{A_{np}} = \frac{2.54 \times 10^{-8} m^2}{(500\ nm)^2} \approx 10^5$ Since each nanopixel stores 3 bits of information, the number of bits stored in one frame is $N_{bits} = 3N_p = 3 \times 10^5$.

**2. The Frame Read-out Time**: a high-speed color camera (e.g. iX Camera i-SPEED 726) is capable of taking images with a resolution of 1064×102 (108528 pixels, enough to image the $10^5$ nanopixels in one frame) at a speed of 200,000 frames per second (FPS). Therefore, the time required to take one color image with high enough resolution is,

$$t_{camera} = \frac{1}{FPS} = \frac{1}{(200000 s^{-1})} = 5 \times 10^{-6} s$$

**3. Time required for the translational stage or disc spinner to move from one frame to another:** In this calculation, we assume a disc spinner with a spinning speed of 5,000 RPM (or 523.6 rad/s) and an APM disc of radius 60 mm (same as a Blu-ray disc). Since the linear velocity varies at different radii, we take an average radius of 30 mm, giving a linear velocity of $v_{lin} = \omega \times R_{disc}$ =532.6 rad/s×30 nm=15.71m/s. Therefore, the time required to move from one frame to another is $t_{spin} = \frac{D_{frame}}{v_{lin}} = \frac{0.18\ mm}{15.71 m/s} = 1.14 \times 10^{-5}$ s. Combining the results of 1, 2 and 3, the data read-out speed of our proposed system is,

$$\rho_{APM} = \frac{N_{bits}}{t_{camera} + t_{spin}} = \frac{3 \times 10^5 bits}{5 \times 10^{-6}s + 1.14 \times 10^{-5}s} = 1.83 \times 10^{10} bits/s$$

or 18.3 Gbits/s. Compare with the maximum data read-out speed of 128 Mbits/s in Blu-ray discs (*37*), and our system shows a read-out speed that is *143 times faster*.

We note that although a frame needs to pass four cameras to have its information retrieved, the time required to read information from one frame is *not* multiplied by four. This multiplication is not required because we implement the parallel data acquisition with four cameras in the read-out setup, which take X + 3 images to read the information from a given APM, with X being the number of frames in the APM (see analysis in main text). As X gets large enough, X + 3 can be well approximated by X, effectively representing the case of taking *a single image per frame*.

### S7. Dielectric Permittivity of Al

During fabrication, we use an unpatterned sample to grow Al under the same conditions as the APM sample and characterize the dielectric permittivity of Al using this sample with a Variable-Angle Spectroscopic Ellipsometer, VASE (J. A. Woollam Co., W-VASE). Figure S6 shows the permittivity data fitted by a dispersive material model with one Drude term and three Lorentz oscillator terms

$$\dot{\varepsilon}(\hbar\omega) = \varepsilon_1 + i\varepsilon_2 = \varepsilon_{1\infty} + \sum_k \frac{A_k}{E_k^2 - (\hbar\omega)^2 - iB_k\hbar\omega} \qquad (S6)$$

For $k^{th}$ oscillator, $A_k$ is the amplitude, $E_k$ is the center energy, and $B_k$ is the broadening of the oscillator. $\hbar\omega$ is the photon energy in eV, $\varepsilon_{1\infty}$ is an additional offset term. In our fitting for the Al permittivity, $\varepsilon_{1\infty} = 1.4842$; the other parameters are listed in Table S1:

| k | A (eV²) | B (eV) | E (eV) |
|---|---|---|---|
| 1* | 141.60 | 0.13445 | 0 |
| 2 | 8.8166 | 0.30195 | 1.5293 |
| 3 | 9.9312 | 0.61558 | 1.7383 |
| 4 | 20.048 | 1.9767 | 2.0991 |

* The Drude term

**Table S1. Drude-Lorentz fitting parameters for Al film.**

This experiment-based Drude-Lorentz model is used in all numerical simulations presented in this work.

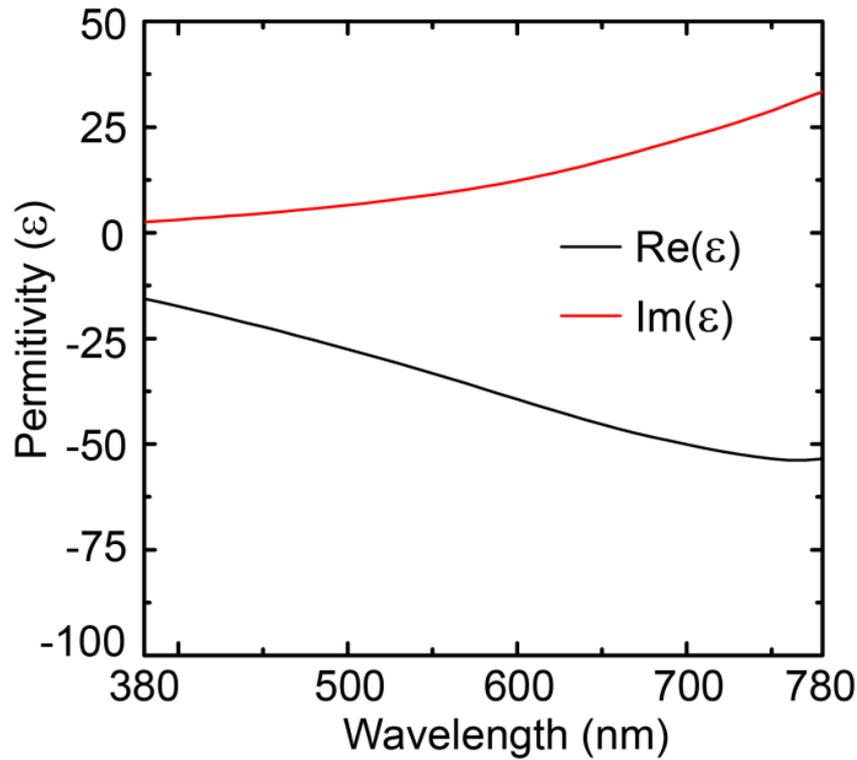

**Figure S6. Dielectric permittivity of Al obtained with VASE measurements.** Real (black) and imaginary (red) parts of the Al dielectric function.

**References**

1. Li, T. *et al.* Manipulating optical rotation in extraordinary transmission by hybrid plasmonic excitations. *Appl. Phys. Lett.* **93**, 021110 (2008).

2. Kim, J. *et al.* Role of epsilon-near-zero substrates in the optical response of plasmonic antennas. *Optica* **3**, 339-346 (2016).